# Zircon survival in shallow asthenosphere and deep lithosphere

Anastassia Y. BORISOVA[1,2*], Ilya N. BINDEMAN[3,4], Mike TOPLIS[5], Nail ZAGRTDENOV [1], Jérémy GUIGNARD [5], Oleg SAFONOV[6], Andrew BYCHKOV[2], Svyatoslav SHCHEKA[7], Oleg E. MELNIK[8], Marion MARCHELLI [1], Jerome FENRENBACH[9]

[1] *Géosciences Environnement Toulouse, GET, Université de Toulouse, CNRS, IRD, UPS, France, 14 Avenue E. Belin, 31400 Toulouse, France (nailzagrtdenov@gmail.com; marionmarchelli@gmail.com)

[2] *Geological Department, Lomonosov Moscow State University, Vorobievu Gory, 119899, Moscow, Russia

[3] *Geological Sciences, University of Oregon, 1275 E 13th street, Eugene, OR, USA

[4] *Fersman Mineralogical Museum, Moscow, Russia

[5]Institut de Recherche en Astrophysique et Planétologie (IRAP) UPS OMP – CNRS - CNES 14 Avenue E. Belin, 31400 Toulouse, France (jeremy.guignard@gmail.com)

[6] *Institute of Experimental Mineralogy, 142432, Chernogolovka, Moscow region, Russia

[7] *Bavarian Research Institute of Experimental Geochemistry and Geophysics (BGI), University of Bayreuth, 95440 Bayreuth, Germany

[8] *Institute of Mechanics, Moscow State University, 1- Michurinskii prosp, 119192, Moscow, Russia

[9] *Institut de Mathématique de Toulouse, Université Paul Sabatier 118, route de Narbonne, 31062, Toulouse, France

*Corresponding author: E-mail: anastassia.borisova@get.omp.eu;

*Corresponding address: Géosciences Environnement Toulouse UMR 5563, Observatoire Midi Pyrénées, 14 Avenue E. Belin, 31400 Toulouse, France; Tel: +33(0)5 61 54 26 31; Fax: +33(0)5 61 33 25 60

# ABSTRACT

Zircon (ZrSiO$_4$) is the most frequently used mineral for dating terrestrial and extraterrestrial rocks. However, the system of zircon in mafic/ultramafic melts has been rarely explored experimentally and most existing models based on the felsic, intermediate and/or synthetic systems are probably not applicable for prediction of zircon survival in terrestrial shallow asthenosphere. In order to determine the zircon stability in such natural systems, we have performed high-temperature experiments of zircon dissolution in natural mid-ocean ridge basaltic (MORB) and synthetic haplobasaltic melts at temperature of 1250 - 1300°C and pressure range from 0.1 MPa to 0.7 GPa coupled with *in situ* electron probe microanalyses of the experimental products at high current.

Taking into account the secondary fluorescence effect in zircon glass pairs during electron microprobe analysis, we have calculated zirconium diffusion coefficient (2.87E-08 cm$^2$/sec) at 1300°C and 0.5 GPa pressure necessary to predict zircon survival in asthenospheric melts of tholeiitic basalt composition. The data imply that typical 100 um zircons dissolve rapidly (~10 h) and congruently upon the reaction with basaltic melt at mantle pressures of 0.2 – 0.7 GPa. We observed incongruent (to crystal ZrO$_2$ and SiO$_2$ in melt) dissolution of zircon in natural mid-ocean ridge basaltic melt at low pressures < 0.2 GPa and in haplobasaltic melt at elevated 0.7 GPa pressure. Our experimental data raise questions about the origin of zircons in mafic and ultramafic rocks, in particular, in shallow oceanic asthenosphere and deep lithosphere, as well as the meaning of the zircon-based ages estimated from the composition of these minerals. Large size zircon megacrysts in kimberlites, peridotites, alkali basalts and other magmas suggest the fast transport and short interaction between zircon and melt. The origin of zircon megacrysts is likely related to metasomatic addition of Zr into mantle as any mantle melting episode should obliterate them.

# INTRODUCTION

Zircon (ZrSiO$_4$) is the most frequently used mineral for dating terrestrial and extraterrestrial rocks (e.g., Valley et al. 2014a) using a variety of the U-Th-Pb isotopes (Schmitt et al., 2014). At the same time, given the remarkable ability of zircon to retain its chemical and isotopic content, the isotopic composition of O and Hf and trace element contents are interpretable in the context of the zircon source and magmatic or/and hydrothermal protolith (e.g., Iizukaa et al., 2015, Valley et al., 2014a); and Ti content is used for the temperature estimates (Watson et al., 2006). The growing evidences for the presence of zircons in such terrestrial mafic and ultramafic rocks as diorites, gabbros and gabbroids, peridotites, kimberlites, carbonatites, dunites, chromitites, garnet-pyroxenites, dolerites (e.g., Bea et al., 2001; Belousova et al., 2002; 2015, Kostitsyn et al., 2009; 2012; 2015; González-Jiménez et al., 2017) and in the meteorite rocks of the Solar system (Ireland & Wlozka, 1992; Humayun et al., 2013, Valley et al., 2014b, Iizukaa et al., 2015; Bellucci et al., 2019) call for the correct interpretation of the isotopic composition of these zircons. Thus, zircon of variable ages up to 3.2 Ga occurs in oceanic mafic and ultramafic rocks. Most researchers are answering the following questions. Why does it occur? What are the proposed mechanisms? Most authors proposes delamination/recycling of continental lithosphere, a direct involvement from fragments of continental lithosphere or an ancient event of fluid involvement followed by the long-lasting storage at mantle conditions. However, to constrain the mechanisms, it is important to get direct experimental data on zircon solubility, dissolution rates and stability field at high temperatures and pressures. This information is contradictory and uncompleted and not yet accepted by most geochronologists.

Starting from the 1980s, several studies have attempted to measure experimentally zircon dissolution rates and to estimate zirconium diffusion in silicate liquids of the predominantly felsic (and intermediate) composition with varying $H_2O$ contents at high temperatures and pressures (Harrison & Watson, 1983; Ellison & Hess, 1986; Baker & Watson, 1988; Baker et al., 2002; LaTourrette et al., 1996; Nakamura & Kushiro, 1988; Mungall et al., 1999; Koepke & Behrens, 2001; Lungstrom 2003; Behrens & Hahn, 2009; Zhang & Xu, 2016). In addition, numerous works have also investigated the conditions of synthetic felsic to ultramafic liquid saturation with zircon and the zircon solubility (Watson, 1982; Harrison & Watson, 1983; Ellison & Hess, 1986; Watson et al., 2006; Rubatto & Hermann, 2007; Burnham & Berry, 2012; Boehnke et al., 2013; Zhang & Xu, 2016; Gervasoni et al., 2016; 2017; Shao et al., 2018; Borisov & Aranovich, 2019).

There are only four experimental studies on lunar basaltic systems (Dickinson & Hess, 1982), synthetic terrestrial basaltic melts (Shao et al., 2018; Borisov & Aranovich, 2019) and dunites (Anfilogov et al., 2015), yet these data do not currently allow us to construct a general model of zircon stability in natural shallow asthenospheric melts of basaltic composition. The system zircon – natural terrestrial basaltic liquid (e.g., equivalent of natural MORB, mid-ocean ridge basalt) has never been studied experimentally, including saturation concentrations and physical-chemical conditions. Nonetheless, large to huge (up to 25 cm) isotopically homogenous zircon crystals and megacrysts are found in carbonatites and these are used as standards for oxygen isotope analyses (UW-MT, Mud Tank, Valley, 2003). These zircon crystals might have been formed at mantle conditions from some type of ultramafic melts. However, the presence of these zircons in the carbonatite rocks is in contradiction to the recent experimental data by Gervasoni et al. (2016; 2017), suggesting that zircon may not be a primary mineral in a low-silica carbonatite melt. These experimental data on zircon stability in carbonatite melts are available only at 0.7 GPa pressure. Thus, the origin and survival of zircons in mafic and ultramafic rocks remain enigmatic.

Therefore, an explanation of the presence of zircon in mafic and ultramafic rocks remains problematic until the appearance of a new model. Moreover, the current interpretation of zircon composition found in mafic and ultramafic rocks is *a priori* false since there is no experimental data on the influence of high temperature-pressure-volatile content conditions on survival of zircon and its trace Pb-U-Th-O-Ti-REE composition survival in mafic and ultramafic systems. These gaps prevent us from correctly interpreting the age of terrestrial and extraterrestrial rocks based on the zircon composition. It is therefore essential to fill these gaps by an experimental approach, combining experimental methods with in situ micro-analytical methods to constrain the conditions of its survival (kinetics) at high temperature.

## METHODS

**Starting materials**

The used large zircon crystals are zircons from the ~730 Ma Mud Tank carbonatites (Crohn and Moore, 1984) provided by Sebastien Meffre (UTAS, Hobart, Australia). A mid-ocean ridge basalt used in our experiments is a typical moderately differentiated (8.2 wt.% of MgO, M = 3.3; G = 2.5, **Supplementary Material**) glassy tholeiitic basalt (number 3786/3) from Knipovich ridge of the Mid Atlantic Ridge sampled by dredging during 38$^{th}$ Research Vessel Adademic Mstislav Keldysh expedition (Sushchevskaya et al., 2000). The MORB glass has been crushed to powder (<100 µm glass size).

**Experimental techniques**

In order to investigate the zircon behavior in the basaltic systems, we performed experimental runs according to the methodologies developed in our current work (e. g., Borisova et al., 2017). We investigated reaction of natural Mud Tank zircon with basaltic liquids of natural mid-ocean ridge basalt and eutectic composition of anorthite-diopside system and this by varying the pressure at similar zircon to basaltic glass ratio (**Table 1**). To study the effect of pressure, similar experiments were performed at varying pressures of 0.1 MPa, to 0.2, 0.5 and 0.7 GPa corresponding to the depths of crustal basaltic chamber (up to 20 km), typical of mafic magma transport conditions in oceanic settings. Seven experimental runs have been performed at 0.1 MPa to 0.7 GPa pressures at temperature range of 1250 - 1300°C with low ratios of zircon to basalt (0.08 to 0.16), (**Table 1, Figs. 1,2**).

The experimental design included a fragment of the double-polished zircon ship in the bottom part and dry or wet basaltic glass powder in the upper part of the $Au_{80}Pd_{20}$ capsule. The masses of the starting components have been weighed. The distilled water$^{MQ}$ has been mixed to the basaltic powder in one experiment (Z7) with additional water (**Table 1**). Because the run duration was much shorter than 48 hours which is the run duration necessary to reach oxygen fugacity equilibrium with a piston-cylinder double-capsule techniques of mineral buffer (Matjuschkin et al., 2015), the redox conditions in our kinetic experiments have been controlled by the initial $Fe^{2+}/Fe^{3+}$ ratios established in the natural samples used as starting materials (MORB glass) as well as those established during the runs due to partial exchange with $Al_2O_3$ pressure media (piston cylinder experiment). The redox conditions established during the shortest runs (<4 h) were estimated to range from QFM(+1.5) to QFM(+4.6).

Two piston-cylinder systems have been used in our experiments. The experiment (Z1) used the "Max Voggenreiter" end-loaded Boyd-England piston-cylinder apparatus at the Bavarian Research Institute of Experimental Geochemistry and Geophysics (BGI), Bayreuth, Germany. Talc cells 3/4 inch in diameter with Pyrex sleeves were used. A tapered graphite furnace was inserted in each cell. Alumina ($Al_2O_3$) spacers were used as pressure-transmitting medium. An $Au_{80}Pd_{20}$ capsule loaded with starting materials was set in the central part of the assembly. A 20% pressure correction was applied for the friction between the talc cell and pressure vessel. A molybdenum disulfide ($MoS_2$) lubricant was introduced to minimize the friction. The temperature in the upper part of the capsules was controlled by a EUROTHERM (2404) controller via either $W_3Re_{97}/W_{25}Re_{75}$ (type D) or $Pt_6Rh_{94}/Pt_{30}Rh_{70}$ (type B) thermocouple accurate to ±0.5°C. The sample was compressed to 0.5 GPa during a period of 20 minutes and then heated up to the run temperature (1300 °C) at a rate of 100 °C/min. The samples were maintained at run conditions during desired durations. The experiments were quenched by switching off electricity. We have applied decompression during periods from 20 minutes to 2 hours. The rate of quenching to the ambient temperature was up to 80°C/min.

Experiments (Z4, Z6, Z7) used the end-loaded Boyd-England piston-cylinder apparatus at the Korzhinskii Institute of Experimental Mineralogy, Chernogolovka, Russia. Standard talc-Pyrex cells 3/4 inch in diameter, equipped with tube graphite heaters and inserts made of MgO ceramics were used as pressure-transmitting medium. The pressure at elevated temperatures was calibrated against two reactions of brucite = periclase + $H_2O$ and albite = jadeite + quartz equilibria. A pressure correction (12%) was introduced for the friction between the cell and hard-alloy vessel. To minimize the friction, a Pb foil and molybdenum disulfide ($MoS_2$) lubricant were used. $Au_{80}Pd_{20}$ capsules with starting mixtures were mounted in the central parts of the cells. The temperature in the upper part of the capsules was controlled to be accurate to ±1°C using a MINITHERM controller via a $W_{95}Re_5/W_{80}Re_{20}$ thermocouple insulated by mullite and $Al_2O_3$ without pressure correction. For the Pyrex-bearing assemblies, the sample was heated to 550 – 600°C at low confining pressure (0.15 – 0.2 GPa) for a few minutes in order to soften the Pyrex glass, subsequently both temperature and pressure were increased almost simultaneously up to the desired run conditions. The samples were maintained at run conditions during desired durations. The experiments were quenched by switching off electricity. The quench rate was 100–200°C/min.

Experiments Z2 – Z3 used a vertical gas-mixing furnace at one atmosphere pressure, 1300°C and controlled oxygen fugacity at the IRAP, Toulouse, France (Toplis et al., 1994). Redox conditions in the furnace were controlled by mixtures of CO and $CO_2$ corresponding to the quartz-fayalite-magnetite (QFM) buffer. The samples were held on Pt wire loops (the wire 0.3 mm in diameter and the loop typically 1 – 2 mm in diameter). For experiments using MORB at the QFM buffer, the wire loops were pre-saturated in iron at the relevant $f_{O2}$ for 24 hours at 1400°C using an Fe-rich basalt powder (Toplis et al., 1994). These wire loops were cleaned using hydrofluoric acid and rinsed with MQ water to get rid of residual salts. Droplets of these mixtures were attached to the wire loops during a 30 s heat treatment at 1400°C in air in a muffle furnace. Once samples of the four different starting mixtures were ready, they were mounted together on a Pt-basket and then introduced into the hot spot of the gas mixing furnace. Samples were quenched by melting the thin suspension wire using an electric current. In this way, the basket falls into the cold zone of the furnace, still under CO-$CO_2$ mixture. The quench rate with this method is estimated at 1000°C/s. After evacuation of the CO, the samples were recovered and mounted in epoxy and polished by SiC grinding paper.

For the experiment Z10, we have used internally heated gas pressure vessel at the Korzhinskii Institute of Experimental Mineralogy, Chernogolovka, Russia. The pressure in the system was created by pure Ar gas. The system was heated by a furnace with two windings (to minimize the thermal gradient). The temperature was set and measured by a TRM-101 OVEN controller through two S-type ($Pt_{90}Rh_{10}$ vs $Pt_{100}$) thermocouples. The thermocouples were mounted at the top and close to the bottom of the run hot spot to monitor the temperature gradient. The duration of experiment was 5 h. The experiment was quenched by switching off the furnace. The pressure during the quench was maintained constant down to 550 °C, and then slowly released. The cooling rate from 1250 to 1000 °C was 167 °C/min, and then 90 °C/min down to 550 °C. After the runs, the capsule was mounted in epoxy, cut in two parts using a diamond saw, and then polished using SiC sand papers and diamond pastes.

**Microanalytical methods**

To study in details the reaction zones using trace element profiles (Zr) (**Fig. 1**) we have used a new generation of electron probe microanalysis (CAMECA SX-Five). Major element analyses of minerals and glasses (**Supplementary Material**) and the sample imaging were performed at the Géosciences Environnement Toulouse (GET, Toulouse, France) laboratory and at the Centre de Microcaractérisation Raimond Castaing (Toulouse, France). The main experimental phases (baddeleyite, zircon and glasses) in the samples have been identified by EDS microprobe technique using a scanning electron microscope (SEM) JEOL JSM-6360 LV with energy-dispersive X-ray spectroscopy (EDS) in GET, Toulouse, France. Major and minor (Zr) elements in the crystals and glasses were analyzed using a CAMECA SX-Five microprobe in Centre Castaing, Toulouse, France. For the electron microprobe CAMECA SX Five, we have applied the following conditions for the electron beam: an accelerating voltage 15 kV, currents of 10 and 20nA for the major elements depending on the resistance of the material under the electron beam, and 100 nA for the minor Zr. To avoid damage on less resistant materials (hydrous glasses), defocused conditions were used up to 10 µm for the beam diameter on the surface of the glass phases. The following synthetic and natural standards were used for calibration: albite (for Na), corundum (Al), wollastonite (Si, Ca), sanidine (K), pyrophanite (Mn, Ti), hematite (Fe), periclase (Mg), $Cr_2O_3$ (Cr), nickel metal (Ni) and reference zircon (Zr). Element and background counting times for most analyzed elements were 5 s for Na and K and 10 s for other major elements, whereas, peak counting times were 120 s for Cr, from 80 to 100 s for Ni and 240 s for Zr. Detection limits for Cr and Ni were 70 ppm and 100 to 300 ppm, respectively and 70 ppm for Zr. The silicate reference materials as well as MPI-DING glasses (ATHO-G, StHs6/80-G, T1-G, KL2-G, ML3B-G, GOR132-G, GOR128-G) were analyzed as unknown samples to monitor the accuracy of the major and trace element analyses. The accuracy estimated on the reference glasses ranges from 0.5 to 3 % (1σ RSD = relative standard deviation), depending on the element contents in the reference glasses. An effect of the secondary fluorescence (Borisova et al., 2018) have been taken into account during Zr analyzes in the silicate glasses analyzed near the zircons in the Z1 sample (**Fig. 1, Supplementary Materials**). The starting MORB glass containing 0.56 wt% $H_2O$ was analyzed according to the method of Bindeman et al. (2012).

**Kinetic modeling**

Zr diffusion coefficient has been calculated following the method of Harrison and Watson (1983). For that, the Zr (ppm) profiles perpendicular to the zircon-glass interface in the Z1 experiment has been measured and the correction has been made based on calculation of the secondary fluorescence effect in the same zircon and basaltic glass (Borisova et al., 2018). The resulting profile has been treated by linear interpolation with the function 'numpy.interp' of Python language.

Kinetic modeling of the bulk dissolution of a spherical zircon was performed using the method of Bindeman and Melnik (2016) with the obtained diffusion coefficient $D_{Zr}$=2.87E-08 $cm^2$/sec at 1300°C and 0.5 GPa (**Supplementary Materials**). The parameters used in the calculations were the interface concentration of Zr in the basaltic melt of 20,840 ppm and the initial Zr concentration of 94 ppm in the starting basaltic melt. Spherical geometry has been applied as that used in the above-mentioned model. The results demonstrate that sphere of zircon of 100 microns in diameter will survive for 9.7 hours, 50 microns – 2.6 hours and 10 microns – 0.2 hours.

## RESULTS

Z1 and Z10 samples are composed of partially dissolved zircon with basaltic glass with linear interface between the zircon crystal and the final glass (**Figs. 1**). At the interface between the both phases, profiles of Zr concentrations have been obtained for the Z1 sample and demonstrated on **Figure 1**; and the interface melt $C_o$ content has been obtained when possible from all experiments (**Table 1**). The mathematic transformation of Zr contents in the profiles to the function of error ($erf^{-1}$) correlates linearly with the distance from the interface with the slope of 0.022, suggesting diffusional control of the zircon dissolution in the basaltic melt according to the criterion described by Harrison & Watson (1983).

Z2 and Z3 samples are made of partially dissolved zircon overgrown by baddeleyite micro-crystals associated with the zone of Si-rich glass (**Fig 1**). Z4 sample contains optically homogeneous basaltic glass, whereas chemical heterogeneity likely due to "vortex"-type convection is remarkable on the BSE image in the lower part of the capsule. The incongruent dissolution has to happen at low-pressure conditions in natural tholeiitic basaltic melt and the conditions of low $SiO_2$ activity in haplobasaltic melt. Indeed, Z6, Z7 samples are composed of the partially dissolved zircon with basaltic glass, where zircons crystals are partially overgrown by baddeleyite micro-crystals associated with the Si-rich glass (**Fig. 1**).

# DISCUSSION

**Zircon versus baddeleyite stability during dissolution**

**Figure 2** demonstrates a diagram of the composition of experimental samples as the result of zircon dissolution at 1250°C to 1300°C with a unique starting configuration of zircon and natural basaltic melt. It is remarkable that at low pressures (0.1 MPa), zircon dissolves in natural basaltic melt incongruently through crystallization of baddeleyite ($ZrO_2$) and a liberation of $SiO_2$-rich melt upon the reaction with natural basaltic MORB liquid. Additionally, baddeleyite is formed at moderate pressures of 0.7 GPa in reaction of zircon with synthetic haplobasaltic (low $a_{SiO2}$) melt, whereas no baddeleyite precipitation happens at elevated pressures (0.2 to 0.7 GPa) upon the zircon dissolution in natural tholeiitic basalt melt, suggesting congruent zircon dissolution at these conditions.

The principal reaction controlling the zircon behavior in the terrestrial systems is the following:

$(ZrSiO_4)_{Zrc} = (ZrO_2)_{Bd} + (SiO_2)_{melt}$ (eq. 1),

where $(ZrSiO_4)_{Zrc}$ is solid zircon crystal, $(ZrO_2)_{Bd}$ – solid baddeleyite and $(SiO_2)_{melt}$ is $SiO_2$ component into the melt. The equation (1) implies that increasing activity of silica shifts the reaction to zircon stability, suggesting that zircon prefers to crystallize in felsic (higher $a_{SiO2}$), rather than mafic and ultramafic systems. In contrast, this equation predicts baddeleyite stability in low $a_{SiO2}$-systems, such as ulramafic (e.g., carbonatite, kimberlite) systems. For example**,** Gervasoni et al. (2017) has investigated the system of zircon in low-$a_{SiO2}$ carbonanite melt at 0.7 GPa and these authors have described stability of baddeleyite and incongruent dissolution of zircon crystals with formation of corona of baddeleyite. In contrast, no baddeleyite crystallization has been observed by Shao et al., (2018) in the synthetic basalt system at pressures ≥ 0.5 GPa, suggesting baddeleyite stability at low pressures in such basaltic systems.

**Zr diffusivity and zircon solubility in silicate melts**

Most existing models considering zircon solubility and dissolution are based on felsic and intermediate and synthetic systems (zircon - felsic and intermediate silicate liquids, e.g., Harrison and Watson, 1983; Boehnke et al., 2013; Zhang & Xu, 2016; and references therein), where zircon is apparently less soluble and more stable compared to the natural mafic and ultramafic systems. The main factors controlling solubility of zircon in the silicate melts are temperature, the melt composition expressed in terms of ratio of nonbridging oxygens per tetrahedrally coordinated cations (NBO/T) which is a simple measure of silicate melt structure based on the ratio between network formers and network modifiers or M (((Na + K + 2 · Ca)/(Al · Si)) (all in cation ratio)), G (where numerator (3 · $Al_2O_3$ + $SiO_2$) represents the network formers, whilst the denominator ($Na_2O$ + $K_2O$ + CaO + MgO + FeO) represent the network modifiers of the melt) or B factor expressed in oxide mole fractions in the melts (0.14($X_{TiO2}/X_{SiO2}$) + 1.3($X_{CaO}/X_{SiO2}$) + 1.5($X_{Na2O}/X_{SiO2}$)−4.5($X_{K2O}/X_{SiO2}$) − 2.7($X_{Al2O3}/X_{SiO2}$)$^2$ + ($X_{MgO}/X_{SiO2}$)$^2$ − 3.7($X_{CaO}/X_{SiO2}$)$^2$ + 75($X_{K2O}/X_{SiO2}$)$^2$) as well as the melt water contents (e.g., Harrison & Watson, 1983; Gervasoni et al., 2016; 2017; Borisov & Aranovich, 2019 and references therein). It is expected that the synthetic (Fe- and Ti-poor) and natural Fe- and Ti-containing basaltic melts would not be similar in respect to zircon solubility. Our natural basaltic liquid (8.2 wt.% of MgO, M = 3.3; G = 2.5, B = 0.15) provides theoretical solubility of 23,078 ppm (or 2.3 wt% of Zr) according to model of Borisov & Aranovich (2019) which is in a very good accord with the experimentally measured solubility of 20,840 ppm. The factors such as M, G and B are indicators of zircon solubility in silicate melts but may be also used for prediction of Zr diffusivity, although the most efficient factor controlling the elemental diffusivity in silicate melts besides temperature remains the melt viscosity determined primarily by the melt $SiO_2$ content (e.g., Mungall, 2002).

To predict zircon survival in natural basaltic systems, zirconium diffusion coefficient has been estimated based on the Z1 experiment (**Fig. 1**) according to the method of Harrison and Watson (1983). Taking into account the secondary fluorescence effect (Borisova et al., 2018), we have calculated zirconium diffusion coefficient (2.87E-08 $cm^2$/sec) at 1300°C and 0.5 GPa necessary to predict zircon survival in asthenospheric melts of tholeiitic basalt composition (M = 3.3; G = 2.5; B = 0.15). The obtained Zr diffusivity is in perfect agreement with data of LaTourrette et al. (1996) on diffusion of Zr in the synthetic haplobasaltic system (**Fig. 3a**). Using the obtained zirconium diffusion coefficient (2.87E-08 $cm^2$/sec) and the model of Bindeman and Melnik (2016), spherical zircon of 100 microns in diameter, will be completely dissolved in 9.7 hours at 1300°C in tholeiitic basaltic liquid, whereas zircon sphere of 50 microns – 2.6 hours and 10 microns – 0.2 hours. The computed dissolution of 1 cm zircon such as Mud Tank, the dissolution occurs in the mantle during ~96000 h (~11 years). Our preliminary data on dissolution experiment of zircon at 0.5 GPa and 1300° C together with other available data on Zr diffusivity suggest a broad increase in diffusion rate of Zr in depolymerized silicate liquids, such as basaltic melts, with low G factor (**Fig. 3b**).

The available data on diffusion of Zr in synthetic melts (**Fig. 3a**) range in the limit of 6 orders of magnitude. Nevertheless, Zhang and Xu (2016) proposed a model allowing evaluation of Zr diffusivity as a function of the melt composition and water contents, whereas volatile (halogen Cl, F) and $CO_2$ contents were not taken into consideration. To compare the experimentally measured Zr diffusivity to those of the calculated according to Zhang and Xu (2016), we have applied their equation 9 to the melt compositions and available experimental data on Zr diffusivity. The calculated values have been compared to the measured values (**Supplementary Material**) as well as to the measured and calculated values from the other experimental works (**Fig. 4**). Our measured diffusivities for the natural tholeiitic basalt and synthetic haplobasalt (LaTourrette et al., 1996) are comparable and approach the line 1:1 on **Figure 4**, suggesting a good fit to the calculations. Additionally, other experimental data except for those of Koepke & Behrens (2001) spread along the line 1:1. Our data on Zr diffusivity imply that the diffusivity in volatile-poor natural basaltic melts may be closely predicted based on the model of Zhang and Xu (2016).

## IMPLICATIONS

Our first experimental data on zircon dissolution and solubility in natural basaltic system further improve our understanding of zircon presence in Earth's and planetary mafic to ultramafic magmas and rocks. The presence of zircon in the mafic/ultramafic systems requires its survival for hours to years and short residence in shallow asthenosphere and lithosphere (**Fig. 5**). For example, complete dissolution of 1 cm zircon, such as Mud Tank, will happen in ~11 years in the mantle. To explain zircon crystallization from tholeiitic basaltic melts at common upper mantle temperatures of 1300°C, Zr concentrations in melts have to attain 20,000 – 30,000 ppm, because of high solubility of zircon in mafic-ultramafic liquids as we demonstrated here one additional time. Thus, the process of zircon crystallization directly from basaltic melt with common Zr concentration of 10s to 100s ppm is nearly impossible. Zircon presence in mafic and ultramafic rocks may be thus related to some other processes that led to Zr enrichment. These processes include the hydrated mantle partial melting (Borisova et al., 2017), continental felsic crust recycling or/and assimilation (Belousova et al., 2015). Its presence may be also an effect of high degree of fractional crystallization of mafic/ultramafic melts in closed systems (e.g., Borisov & Aranovich, 2019), total zircon dissolution and reprecipitation mechanism (Bea et al., 2001) or/and a metasomatic Si-F-rich fluid contribution (e.g., Louvel et al., 2013). For example, zircon crystals and megacrysts from the Mud Tank Carbonatite complex were suggested to have metasomatic origin due to the mantle-derived magma intrusion at mid-crustal depths in the continental settings (Currie et al., 1992). The Mud Tank zircon-hosted inclusions comprising apatite, magnetite, carbonate and microscopic REE-rich phases are present but not abundant (Gain et al., 2019), suggesting metasomatic rather than magmatic origin of the zircon megacrysts.

The paradox of zircon presence in mafic and ultramafic rocks implying its stability and low solubility in mafic and ultramafic melt contrast with experimental data, suggesting that zircon is highly soluble, rapidly dissolved by diffusion-controlled mechanism. The main mechanism of zircon crystallization at mantle conditions may be participation of a fluid and/or a creation of felsic systems by fractional crystallization, partial melting or fluid-rock interaction (mantle metasomatism). For example, Mud Tank zircons, isotopically homogeneous megacrysts used as isotopic reference material (Valley, 1998; Gain et al., 2019), might be created by such metasomatism. The 731.0 Ma Mud Tank zircons have low contents of REE, U and Th, similar to zircon megacrysts in many kimberlites (Mitchell, 1986). Nevertheless, it is likely impossible to create zircon in the mantle, in equilibrium with mafic-ultramafic rocks. How does it form? Diffusion in zircons of most elements is extremely sluggish (Cherniak, 2010), the isotopic homogenization cannot happen, unless the available diffusion coefficients in zircon are correct (Bindeman et al., 2018).

It should be noted that the primitive mantle-normalized pattern of the primary carbonatite melts (Walter et al., 2008) has no Zr and Hf anomalies compared to the well-known strong Zr and Hf depletions in composition of the carbonatite rock and glasses from the oceanic settings (e.g., Hauri et al., 1993). This fact implies that a magmatic process (e.g., liquid immiscibility or/and fractional crystallization of a mineral phase) might have been responsible for the formation of the strong negative Zr and Hf anomalies at high pressures (perhaps, at ~9 GPa), due to differentiation of the primary carbonatite melts. Perhaps, zircon, reidite or other Zr-rich mineral phase might be such Zr-rich phase responsible for the Zr and Hf depletions and might be formed from the carbonatite melts at deep asthenospheric conditions (> 0.7 GPa). However, zircon stability and solubility have never been investigated experimentally yet at these high-pressure and ultrahigh-pressure conditions.

## CONCLUSIONS

Our data suggest high solubility of zircon in natural basaltic melt and very fast congruent dissolution of zircon in natural basaltic melt at pressures of 0.2 to 0.7 GPa. We have calculated timescales of zircon survival in natural mafic liquids of tholeiitic basalt composition (M = 3.3; G = 2.5; B = 0.15). For example, spherical zircon of 1 cm in diameter will be completely dissolved at 1300°C and 0.5 GPa in tholeiitic basaltic liquid in ~11 years, 100 microns zircon – in 9.7 hours, whereas zircon sphere of 50 microns – 2.6 hours and 10 microns – 0.2 hours. In contrast, incongruent dissolution of zircon at low pressures (< 0.2 GPa) may hamper the bulk zircon dissolution in the tholeiitic basaltic melt because of formation of baddeleyite halo associated with Si-rich melt.

To date correctly the terrestrial and extraterrestrial rocks, a preliminary investigation of the zircon-hosted fluid or/and melt micro-inclusions has to be performed to check any possibility of the zircon crystallization from the mafic or/and ultramafic liquids or/and associated fluids. Our data suggest that zircons have to be dissolved rather than crystallized in contact with natural tholeiitic basaltic (shallow asthenosphere-derived) melts.

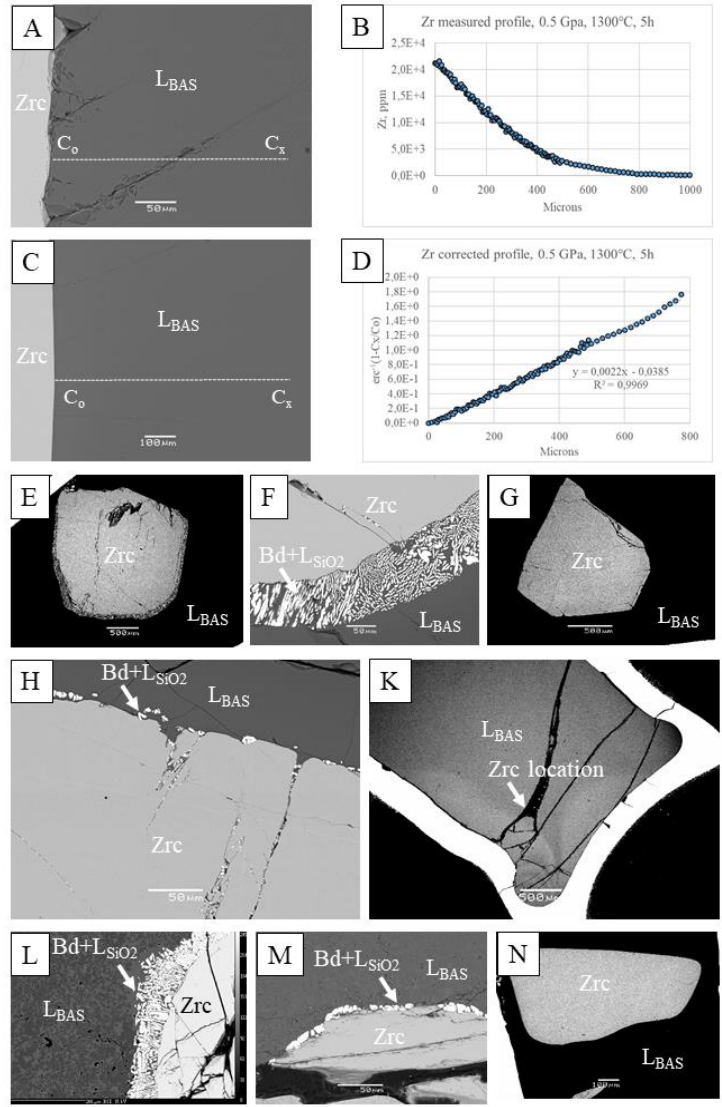

**Figure 1. (A,B,C,D)** Backscattered electron image of (**A**) Z1 and (**C**) Z10 samples demonstrating congruent dissolution of zircon (Zrc) in natural basaltic melt. (**B,D**) The measured and corrected Zr profiles and the function of erf$^{-1}$ (1-$C_x$/$C_o$) suggest that the zircon dissolution is controlled by diffusion in the basaltic melt. (**E,F**) Backscattered electron image of Z2 sample illustrating incongruent zircon dissolution in natural basaltic melt. A reaction with basaltic melt causes zircon replacement by baddeleyite and melt rich in Si. (**G,H**) Backscattered electron image of Z3 sample demonstrating incongruent zircon dissolution in natural basaltic melt. A reaction with basaltic melt causes zircon replacement by baddeleyite (Bd) and liberation of melt rich in Si ($L_{SiO2}$). (**K**) Z4 sample: the bulk zircon dissolution in natural basaltic melt according to the congruent mechanism. (**L**) Backscattered electron image of Z6 sample illustrating incongruent zircon dissolution in natural basaltic melt. A reaction with basaltic melt causes zircon replacement by baddeleyite and liberation of melt rich in Si. (**M**) Backscattered electron image of Z7 sample illustrating incongruent zircon dissolution in natural basaltic melt. A reaction with basaltic melt causes zircon replacement by baddeleyite and melt rich in Si. (**N**) Backscattered electron image of Z10 sample demonstrating congruent zircon dissolution in natural basaltic melt.

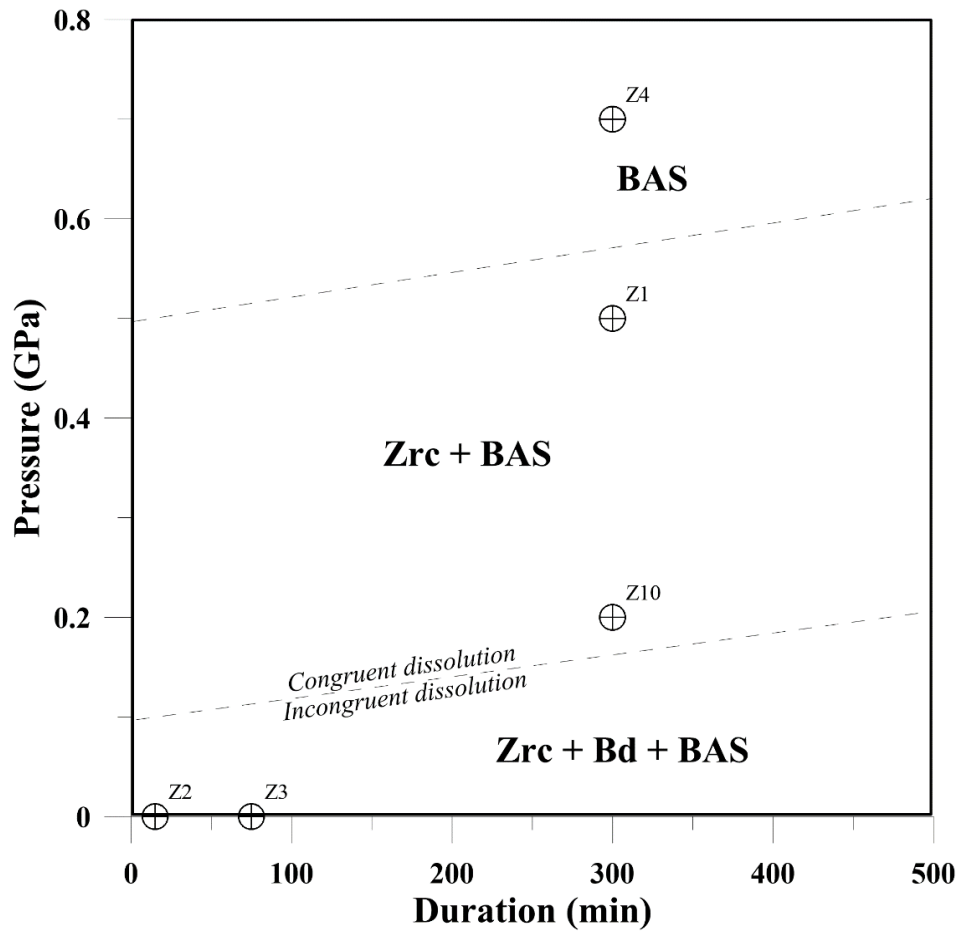

**Figure 2.** Diagram of the composition of volatile-poor experimental samples after the runs on Mud Tank zircon-natural MORB-type basaltic melt pairs vs. the run duration. The experiments 1250°C to 1300°C used different techniques to attain different pressures. BAS – homogeneous basaltic glass; Zrc + BAS – zircon with the reacted basaltic glass; Zrc + Bd + BAS – zircon, baddeleyite with associated Si-rich glass and the reactional basaltic glass as the result of the reaction.

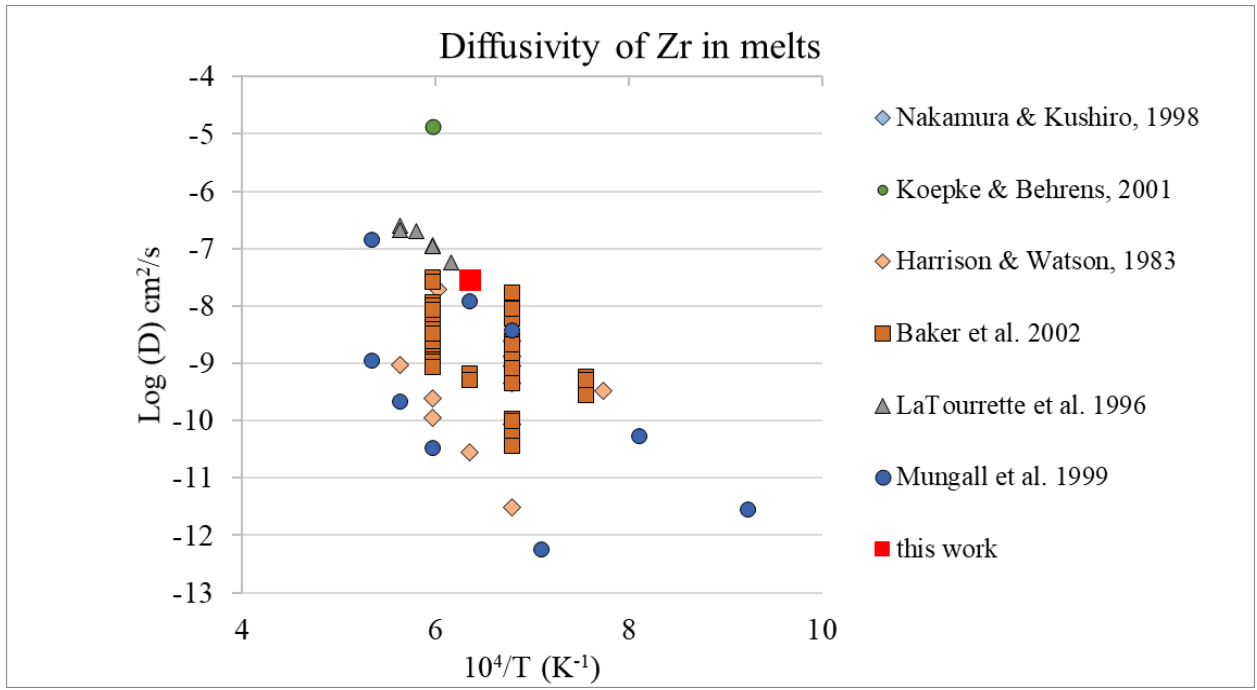

**Figure 3(a).** All available and new (this work) experimental data on Zr diffusivity in dry and hydrous silicate melts (in log(D), cm$^2$/s) vs. temperature (10000/T in K$^{-1}$).

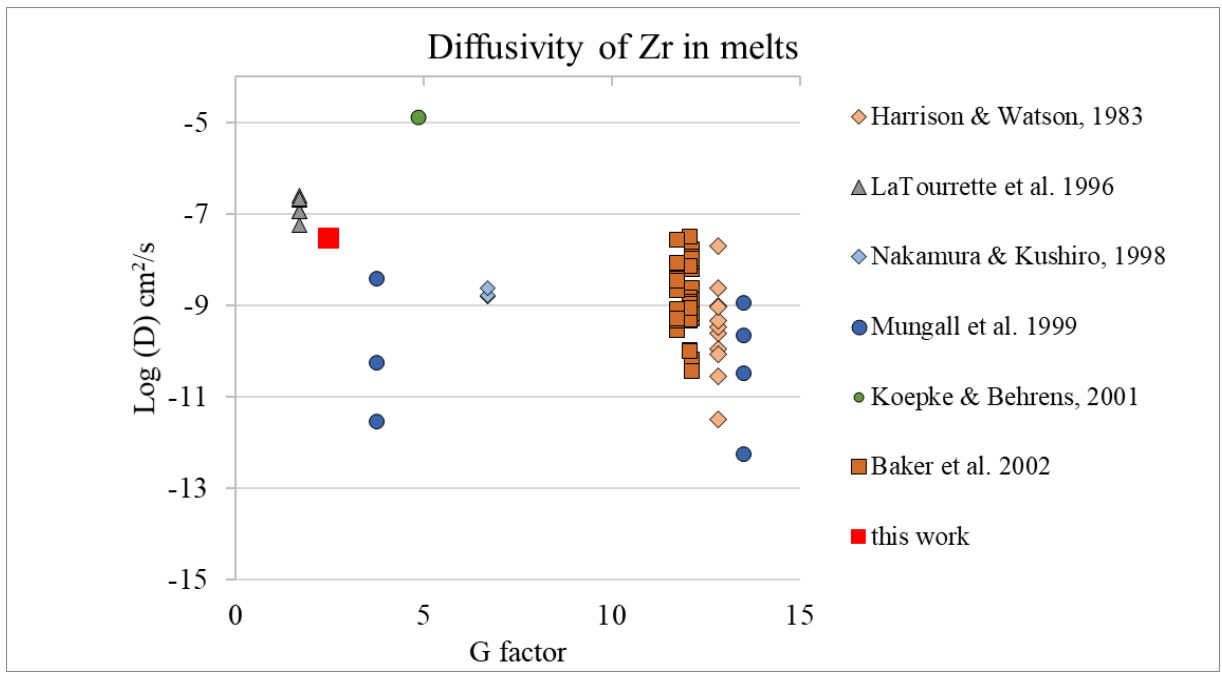

**Figure 3(b).** All available and new (this work) data on Zr diffusivity in dry and hydrous silicate melts (in log(D), cm$^2$/s) vs. G factor of the investigated melts (calculated on the anhydrous basis).

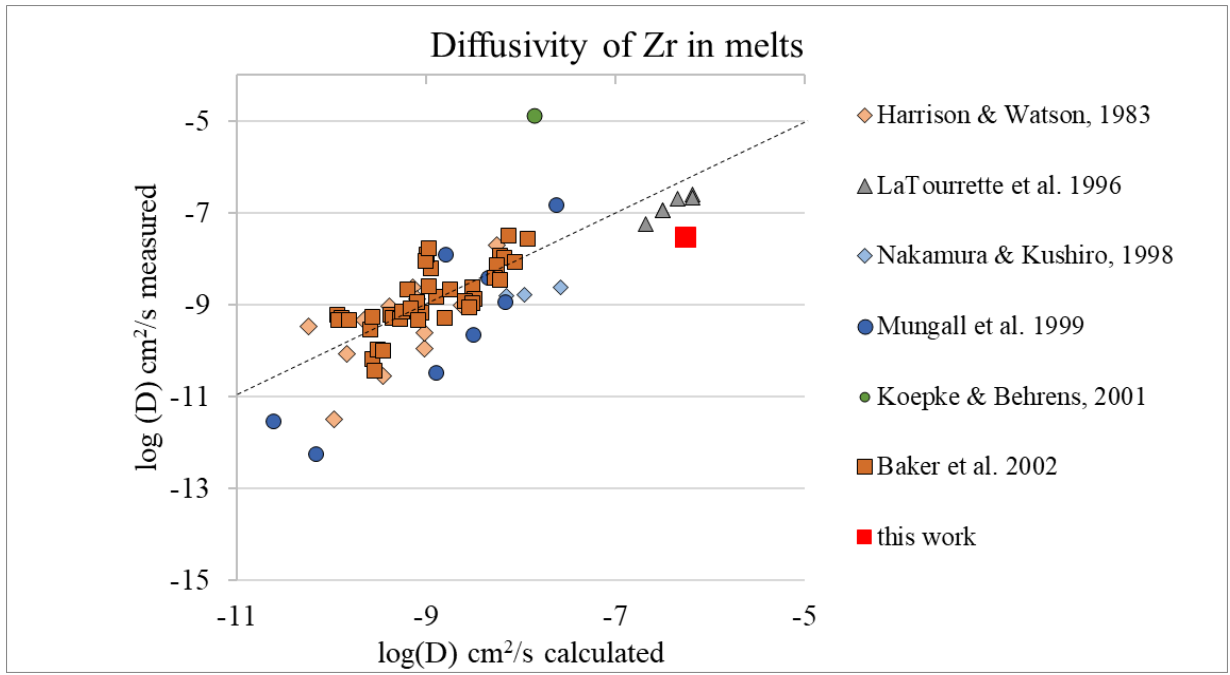

**Figure 4.** All available and new data on the measured Zr diffusivity in dry and hydrous silicate melts (in log(D), cm$^2$/s) vs. calculated Zr diffusivity (in log(D), cm$^2$/s) after Zhang & Xu (2016) (calculated on hydrous basis).

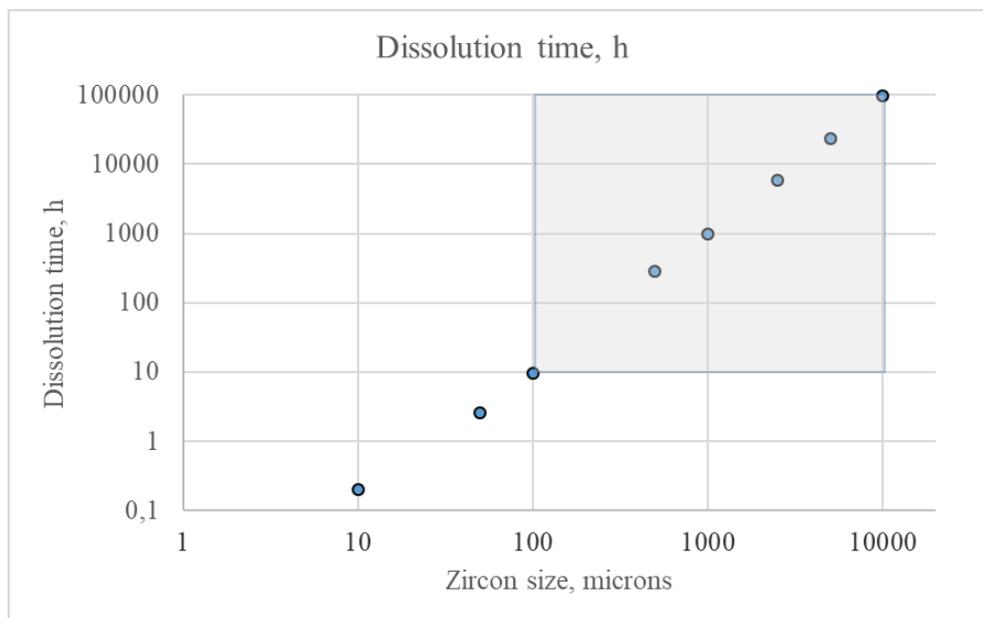

**Figure 5.** The calculated time of zircon dissolution in tholeiitic basaltic melt at 1300°C according to the kinetic model of Bindeman & Melnik (2016). The shaded field corresponds to the dissolution timescales from hours to years according to the typical sizes of zircon crystals and megacrysts in carbonatites and kimberlites.

**Table 1.** Pressure-temperature conditions of experiments on zircon dissolution in natural basaltic and synthetic haplobasaltic melt

| Sample number | Techniques[*] | Conditions (pressure, temperature) | Duration (hours) | Type of basalt & initial mass of Zrc to basalt to water, | Observed final phases[***] | Interface melt $C_o$ content (ppm)[**] |
|---|---|---|---|---|---|---|
| Z1 | BGI, PC | 1300°C, 0.5 GPa | 5h | Zrc to BAS 0.075 | Partially dissolved zircon (PDZ) Zrc, $L_{bas}$ | 20 840 |
| Z2 | IRAP, G | 1300°C, 0.1 MPa, FMQ | 15 min | Zrc to BAS 0.142 | PDZ, Bd, $L_{bas}$ | 27 406 |
| Z3 | IRAP, G | 1300°C, 0.1 MPa, FMQ | 1h 15min | Zrc to BAS 0.086 | PDZ, Bd, $L_{bas}$ | 26 910 |
| Z4 | KIEM, PC | 1300°C, 0.7 GPa | 5h | BAS 0.037 | $L_{bas}$ | - |
| Z6 | KIEM, PC | 1300°C, 0.7 GPa | 5h | Zrc to AnDi 0.157 | PDZ, Bd, $L_{bas}$ | N.A. |
| Z7 | KIEM, PC | 1300°C, 0.7 GPa | 5h | Zrc to AnDi 0.097; $H_2O$ = 7.0 wt% | PDZ, Bd, $L_{bas}$ | N.A. |
| Z10 | KIEM, IHPV | 1250°C, 0.2 GPa | 5h | Zrc to BAS 0.115 | PDZ, $L_{bas}$ | 23 889 |

[*] BGI - Bavarian Research Institute of Experimental Geochemistry and Geophysics (BGI), Bayreuth, Germany; IRAP - Institut de Recherche en Astrophysique et Planétologie (IRAP), Toulouse, France; KIEM - Korzhinskii Institute of Experimental Mineralogy, Chernogolovka, Russia. PS – piston cylinder; G - gas-mixing furnace at one atmosphere pressure; IHPV – internally heated pressure vessel.

[**] - Interface melt concentration in ppm ($C_o$) calculated taking into account the secondary fluorescence effect only for experiments with partially dissolved zircon (PDZ). N.A. – not analyzed.

[***]BAS – natural basaltic glass; AnDi – haplobasaltic glass; FMQ- fayalite-magnetite-quartz redox buffer; PDZ – partially dissolved zircon; Bd – baddeleyite; $L_{bas}$ – basaltic glass.